\documentclass[onecolumn,amsmath,amssymb,nofootinbib,floatfix,society=aps,journal=prd,10pt]{revtex4-2}

\usepackage{graphicx}
\usepackage{subfigure}
\usepackage{overpic}
\usepackage{tikz}

\usepackage{bm}
\usepackage{dcolumn}

\usepackage{color}
\usepackage[pagewise]{lineno}

\usepackage{hyperref}

\makeatletter

\newcommand{\Rmnum}[1]{\expandafter\@slowromancap\romannumeral #1@}
\makeatother

\def\be{\begin{equation}}
	\def\ee{\end{equation}}
\def\bea{\begin{eqnarray}}
	\def\eea{\end{eqnarray}}

\begin{document}
\title{ Violation of cosmic censorship in Einstein-Maxwell-Scalar models with fractional coupling }

\author{Yan-Qing Xu }
\affiliation{Physics Department, College of Physics and Optoelectronic Engineering, Jinan University, Guangzhou 510632, China}
\author{Rui-Feng Zheng}
\affiliation{Physics Department, College of Physics and Optoelectronic Engineering, Jinan University, Guangzhou 510632, China}
\author{Yu-Peng Zhang}
\affiliation{Lanzhou Center for Theoretical Physics, Key Laboratory for Quantum Theory and Applications of the Ministry of Education, Key Laboratory of Theoretical Physics of Gansu Province, School of Physical Science and Technology, Lanzhou University, Lanzhou 730000, China}
\affiliation{Institute of Theoretical Physics ~ Research Center of Gravitation, School of Physical Science and Technology, Lanzhou University, Lanzhou 730000, China}
\author{Cheng-Yong Zhang}
\email[Corresponding author ]{zhangcy@email.jnu.edu.cn}
\affiliation{Physics Department, College of Physics and Optoelectronic Engineering, Jinan University, Guangzhou 510632, China}

\begin{abstract}
The weak cosmic censorship conjecture plays a foundational role in classical gravity by asserting that spacetime singularities are generically hidden behind event horizons. In this work, we explore its robustness in the Einstein–Maxwell–Scalar theory with fractional coupling by studying both static black hole solutions and their fully nonlinear dynamical evolution. We identify a class of scalarized black holes that develop negative energy density near the event horizon, indicating violations of the classical energy conditions. Numerical evolutions of perturbed configurations reveal that sufficiently strong fractional coupling drives rapid curvature growth and geometric degeneration in the near-horizon region, accompanied by persistent negative energy density. While the simulations do not resolve the ultimate end state, the observed dynamics consistently point toward a weakening of the horizon-supporting structure and are suggestive of incipient naked singularity formation. These results uncover a classical mechanism through which fractional coupling can challenge the validity of the weak cosmic censorship conjecture in asymptotically flat spacetimes.
\end{abstract}

\maketitle	

\section{INTRODUCTION}\label{S1}

One of the most profound predictions of general relativity  is that gravitational collapse can lead to the formation of spacetime singularities. The singularity theorems established by Penrose and Hawking \citep{PhysRevLett.14.57,hawking:penrose1970} demonstrate that under certain general physical conditions, regions of divergent curvature inevitably emerge in spacetime. In order to maintain the predictability of classical theories, Penrose further proposed the weak cosmic censorship conjecture (WCCC) \citep{Penrose:1969pc}, this conjecture posits that such singularities are always enclosed by event horizons, rendering them causally disconnected from the external universe. Acting as a  “cosmic veil", the event horizon shields external observers from the unpredictability of singularities, preserves the deterministic nature of classical physics outside the horizon, and ensures the well-defined thermodynamics of black holes. Nevertheless, since its introduction, the WCCC has lacked a rigorous proof, and its universality remains one of the most fundamental and unresolved issues in theoretical physics \cite{wald1997gravitationalcollapsecosmiccensorship,ed3deaa2-9a0c-326e-80db-65101a79480c,Ong:2020xwv,He:2019mqy,Mishra:2019jsr,Hod:2019wcw,Meng:2023vkg,Zhao:2024lts,Lin:2024deg,Cui:2024xqg,Wu:2024ucf,Anand:2024wxa,Corelli_2021}.

Testing the robustness of the WCCC necessarily involves probing its potential mechanisms of violation. This has given rise to two main research directions: The first approach is within the general
relativity framework, employing gedanken experiments, such as attempts to overcharge or overspin a black hole, to challenge the stability of event horizons. The classical work of  Wald demonstrated that an extremal Kerr–Newman black hole cannot be “destroyed” by absorbing charged or spinning test particles \citep{Wald1974,Sorce:2017dst}, thereby supporting the WCCC. Building upon this foundation, subsequent studies have extended the testing of this conjecture to more complex physical scenarios, such as non-extremal black holes and higher-order spins \citep{Jacobson:2009kt,Barausse:2010ka,Siahaan:2015ljs,Liu:2020cji,Zhao:2024qzg,Meng:2024els}. The second approach considers modified theories of gravity \citep{Corelli:2022pio,Corelli:2022phw,Yu:2018eqq,Zeng:2019hux,Liang:2018wzd,Tavakoli:2020xgc}, with the goal of exploring whether naked singularities can emerge either as classical solutions or as dynamical end-states. This perspective is motivated by the broader question of whether the cosmic censorship conjecture is a fundamental property of gravity or a consequence specific to general
relativity. Consequently, the study of naked singularities within modified gravitational frameworks provides a natural arena for testing the robustness of cosmic censorship beyond general relativity.

Among the many possibilities, the Einstein–Maxwell–Scalar (EMS) theory stands out as a particularly useful model, combining relative technical simplicity with a rich phenomenology \citep{Herdeiro_2018,Zhang:2021nnn,Astefanesei_2019,PhysRevD.100.084045,Fernandes_2019,Luo:2022roz,Herdeiro_2020,PhysRevD.103.044019,PhysRevD.100.044015,Myung:2018vug,Zhang:2015jda,Zhang:2021etr,Zhang:2021edm,Zhang:2021ybj,Brihaye:2019gla,Xiong:2022ozw,Zhang:2022cmu,jiang2023typeicriticaldynamical,Chen:2023iws,Zhang_2025,jiang2023typeicriticaldynamical,Zhang:2024wci,Zhang:2025lhm}. The inclusion of a non-minimal coupling between the scalar and electromagnetic fields substantially enlarges the space of classical black hole solutions \citep{Fernandes_2019,Astefanesei_2019,PhysRevD.100.084045}, making the EMS framework especially suitable for investigating violations of cosmic censorship. In particular, fractional coupling forms (e.g., or other rational functions) play a critical role \citep{Fernandes_2019}. Such couplings can induce spontaneous scalarization under specific parameter conditions: when the black hole charge-to-mass ratio exceeds a critical threshold, the scalar-free Reissner-Nordström (RN) black hole solution becomes linearly unstable, and new solutions with non-trivial scalar field configurations—scalarized black holes, emerge as stable ground states.  
It is worth emphasizing that under certain fractional couplings, scalarized black hole solutions may exhibit negative energy density near the event horizon \citep{Fernandes_2019}. This phenomenon directly violates the classical null energy condition, whose breakdown may not only hinder the formation and maintenance of the horizon but also provide a potential pathway for the emergence of naked singularities. The anomalous energy distribution induced by non-minimal coupling generates an effective gravitational repulsion, thereby counteracting the focusing effect required to sustain the horizon. This raises the intriguing possibility that scalarized black holes in EMS theories could dynamically evolve toward naked singularities, providing a classical mechanism for violating the WCCC.

In this work, we investigate this possibility by studying the static and dynamical properties of black hole solutions in EMS models with fractional coupling functions.  We first construct static black hole solutions and map their phase diagram, identifying a critical region where hairy and hairless solutions coexist. In this regime, the hairy branch exhibits negative energy density near the horizon—a clear pathology threatening horizon stability. We then probe the dynamical implications of this configuration by simulating the nonlinear evolution of perturbed black holes. Our numerical results demonstrate that, within finite time, the Kretschmann scalar near the horizon tends toward divergence, indicating the disruption of the horizon and the dynamical formation of a naked singularity. We argue that the
underlying physical mechanism is intrinsically related to the negative energy density near the horizon.  The effective repulsive force generated by this negative energy density may ultimately “tear” the horizon, exposing the singularity
to external spacetime. This study not only proposes a dynamical mechanism for WCCC violation linked to fractional couplings but also highlights the pivotal role of energy conditions in determining the ultimate fate of gravitational collapse.

The paper is structured as follows: In Sect. \ref{S2}, we introduce the EMS model and fractional coupling functions, solve the equations of motion in PG coordinates, construct the solution phase diagram and analyze the static solutions, revealing the existence of regions with negative energy density; In Sec. \ref{S3}, by applying small perturbations to the black hole in the strange region, we found evidence of the existence of a naked singularity during the dynamic numerical evolution process; And in Sec. \ref{S4}, we summarized the conclusions and discussed future research directions. It should be noted that in this paper, we adopt natural units.
 
 \section{The EMS Model }\label{S2}
 
The action for EMS theory is given by
 \begin{equation}\label{eq1}
S = \frac{1}{16\pi} \int d^{4}x \sqrt{-g} \left[ R - 2\nabla_{\mu}\phi \nabla^{\mu}\phi - f(\phi) F_{\mu\nu} F^{\mu\nu} \right]~,
\end{equation}
where $R$ is the Ricci scalar, $F_{\mu\nu}=\partial_{\mu}A_{\nu}-\partial_{\nu}A_{\mu}$ is the electromagnetic field tensor, 
$A_{\mu}$ describes the gauge potential of the charge carried by the black hole, and $\phi$ is a real scalar field which non-minimally couples to the Maxwell term via the coupling function $f(\phi)$.
By varying Eq. (\ref{eq1}), we obtain the Einstein field equations
$$R_{\mu\nu} - \frac{1}{2}R g_{\mu\nu} =2(T^{\phi}_{\mu \nu}+f(\phi)T^A_{\mu \nu})~,$$
 the scalar field equation
$$\nabla_{\mu}\nabla^{\mu}\phi = \frac{1}{4}\frac{df(\phi)}{d\phi}F_{\mu\nu}F^{\mu\nu}~,$$ 
and the Maxwell equations 
$$\nabla_{\mu}\left(f(\phi) F^{\mu\nu}\right) = 0~,$$ 
where the energy-momentum tensor $ T^{\phi}_{\mu \nu}=\partial_{\mu}\phi \partial_{\nu}\phi - \frac{1}{2}g_{\mu\nu} \nabla_{\rho}\phi \nabla^{\rho}\phi$ and $T^A_{\mu \nu}=F_{\mu\rho} F_{\nu}^{\ \rho} - \frac{1}{4} g_{\mu\nu} F_{\rho\sigma} F^{\rho\sigma}$. Here, we adopt the following form of fractional coupling
\begin{equation}\label{eq2}
f(\phi) = \frac{1}{1 + b \phi^{2}}~,
\end{equation}
where $b$ is the coupling parameter that quantifies the coupling strength. 
Fractional coupling scenarios have revealed a variety of intriguing physical implications in recent studies. They have been extensively employed to investigate the triggering mechanisms of spontaneous scalarization in black holes \citep{Fernandes_2019,Astefanesei_2019,PhysRevD.100.084045,Luo:2022roz,Herdeiro_2020,PhysRevD.103.044019,PhysRevD.100.044015}, as well as to analyze horizon-scale phase transition phenomena \citep{Gubser_2005}.
 It is worth noting that the fractional coupling suggests  dynamical formation of naked singularities in asymptotically AdS spacetimes \citep{Luo:2022roz}, where the reflective AdS boundary and the confining nature of the background geometry play a crucial role in nonlinear evolution.
In contrast, whether similar violations of the cosmic censorship conjecture can occur in asymptotically flat spacetimes, where energy can disperse to infinity and no background confinement is present, remains largely unexplored. Addressing this question is essential for clarifying whether the emergence of naked singularities is an intrinsic consequence of the local energy condition violations induced by fractional coupling, or merely a feature tied to the global structure of AdS spacetime.

\subsection{ Equations of Motion }
To deal with the dynamical simulation of the black hole, we adopt the spherically symmetric Painlevé-Gullstrand (PG) coordinate system,
\begin{equation}\label{eq3}
ds^2=-(1-\zeta^2)\alpha^2dt^2+2\zeta\alpha dtdr+dr^2+r^2(d\theta^2+\sin^2\theta d\phi).
\end{equation}
Here $\alpha$ is the lapse function and $\zeta$ the shift function, both of which are dependent on $t, r$, with $\zeta(t, r) = 1$ marking the location of the apparent horizon. 
For the EMS model with fractional coupling, both hairy (scalarized) and hairless (bald) black hole solutions exist. The hairless RN solution is recovered when $\alpha=1$ and $\zeta = \sqrt{\frac{2M}{r} - \frac{Q^{2}}{r^{2}}}$, where $M$ denotes the total mass of the system and $Q$ represents the electric charge of the black hole. 

The gauge field is chosen as $A_\mu dx^\mu=A(t,r)dt$. Substituting into the Maxwell field equations, we obtain
\begin{equation}\label{eq4}
\partial_r A = \frac{Q\alpha(1+b \phi^2 )}{r^2}.
\end{equation} 
For the purpose of numerically simulating Einstein field equations, scalar field equation and Maxwell  field equations, we introduce two auxiliary variables
\begin{align}
\Pi &=\frac{1}{\alpha}\partial_t\phi-\zeta \Phi~,\label{eq121}\\
\Phi &=\partial_r \phi.\label{eq122}
\end{align}
Substituting them into the Einstein equations and simplifying, we obtain the constraint equations for $\alpha$ and $\zeta$, as well as the time evolution equation for  $\zeta$
\begin{align}
\partial_r \alpha &= -\frac{r\Pi\Phi\alpha}{\zeta}~,\label{eq131} \\
\partial_r \zeta &= \frac{r}{2\zeta}\left(\Phi^{2} + \Pi^2\right) + \frac{Q^2(1+b\phi^2)}{2r^3\zeta} + r\Pi\Phi - \frac{\zeta}{2r}~,\label{eq132}\\
\partial_t \zeta &= \frac{r\alpha}{\zeta}\left(\Pi + \Phi\zeta\right)\left(\Pi\zeta + \Phi\right)~.\label{eq133}
\end{align}
Similarly, substituting $\Pi$ and $\Phi$ into scalar field equation and simplifying, we obtain the following time evolution equation for the scalar field  as
\begin{align}
\partial_t \phi &= \alpha(\Pi + \Phi\zeta)~,\label{eq141} \\
\partial_t \Pi &= \frac{\partial_r\left[(\Pi\zeta + \Phi)\alpha r^2\right]}{r^2} + \frac{b \phi \alpha Q^2}{r^4 }~, \label{eq142}\\
\partial_t \Phi &= \partial_r\left[\alpha(\Pi + \Phi\zeta)\right]~.\label{eq143}
\end{align}
Equations \eqref{eq141}-\eqref{eq143} characterize the dynamical evolution of the system. If we further consider the case where the system reaches an equilibirum state (i.e., all physical quantities become time-independent), a set of simplified relations can be derived from \eqref{eq141}-\eqref{eq143}. Substituting these relations into the constraint equations  \eqref{eq131}-\eqref{eq133} then yields the static field equations, 
\begin{align}
0 &= \partial_r\alpha - r\alpha\left(\partial_{r}\phi\right)^2~, \label{eq:alpha} \\
0 &= \partial_r\zeta + \frac{\zeta}{2r} - \frac{r\left(1 - \zeta^2\right)}{2\zeta}\left(\partial_r\phi\right)^2 - \frac{Q^2(1+b \phi^2)}{2r^3\zeta }~, \label{eq:zeta} \\
0 &= \partial_r^2\phi + \frac{1}{\left(\zeta^2-1\right)}
    \left[
        \left(\frac{Q^{2}(1+b\phi^2)}{r^2} + \zeta^2 - 2\right)
        \frac{\partial_r\phi}{r} + \frac{b\phi Q^2}{r^4}    
    \right]~.\label{eq:phi}
\end{align}
Equations  \eqref{eq:alpha}-\eqref{eq:zeta} are a set of nonlinear coupled field equations satisfied by the gravitational fields 
$(\alpha,\zeta)$ and the scalar field $\phi$ in the static solutions. 

\subsection{Relevant Physical Quantities}

In order to analyze the global structure of the static solutions and to investigate possible signatures of spacetime singularities, we introduce several physically relevant quantities. These include the Misner--Sharp (MS) mass, which characterizes the quasi-local energy distribution, the Kretschmann scalar, providing a coordinate-independent measure of spacetime curvature, and the energy density, which serves as a diagnostic of the energy conditions.

Within general relativity, the MS mass $m(t,r)$ is one of the most widely used quasi-local mass definitions. In the coordinate system adopted here, it takes the form
\begin{equation}\label{eq8}
m(t,r)=\frac{r}{2}\left(1-g^{\mu\nu}\partial_{\mu}r\,\partial_{\nu}r\right)
      =\frac{r}{2}\,\zeta(t,r)^2 .
\end{equation}
At spatial infinity, the MS mass approaches the total mass of the system,
\begin{equation}\label{eq:M}
M=\lim_{r\to\infty}m(t,r)
 =\lim_{r\to\infty}\frac{r}{2}\,\zeta(t,r)^2 .
\end{equation}
It is worth noting that in PG coordinates the ADM mass identically vanishes and therefore fails to capture the global mass of the spacetime. Consequently, throughout this work we adopt the MS mass $m(t,r)$ as the appropriate mass measure~\cite{doi:10.1142/9692}.

In order to probe potential curvature singularities, we compute the Kretschmann scalar invariant
$K=R^{\alpha\beta\gamma\delta}R_{\alpha\beta\gamma\delta}$.
In the present framework, it can be expressed explicitly as
\begin{equation}\label{eq10}
K=8(\Pi^2-\Phi^2)\!\left[\Pi^2-\Phi^2-\frac{8\zeta^2}{r^2}\right]
 +\frac{12\zeta^4}{r^4}
 +\frac{4Q^2(1+b\phi^2)}{r^4}
 \left[\frac{5Q^2(1+b\phi^2)}{r^4}
       +4(\Pi^2-\Phi^2)
       -\frac{6\zeta^2}{r^2}\right].
\end{equation}
For the conventional RN solution, the Kretschmann scalar behaves as
$K\sim M^2/r^6+\mathcal{O}(M^3/r^7)$, whose leading divergence as $r\to0$ signals the presence of a central curvature singularity. Deviations from this behavior in Eq.~\eqref{eq10} therefore provide important information on the singular structure of the present model.

The formation of spacetime singularities is closely related to the behavior of the energy conditions. The energy density measured by a timelike observer whose four-velocity coincides with the hypersurface normal
$n_\mu=(-\alpha,0,0,0)$ is given by
\begin{equation}\label{eq11}
\rho(r)=n^{\mu}\!\left(T^{\phi}_{\mu\nu}
+f(\phi)T^{A}_{\mu\nu}\right)n^{\nu}
=\frac{1}{2}\left(\Pi^2+\Phi^2\right)
 +\frac{Q^2(1+b\phi^2)}{2r^4}.
\end{equation}
The weak energy condition requires $\rho(r)\ge0$. By examining the spatial distribution and temporal evolution of $\rho(r)$, in particular its sign, one can assess whether the weak energy condition is violated. Such violations are commonly regarded as important indicators of pathological spacetime behavior and may signal the onset of singularity formation.

Together, these physical quantities provide a consistent framework for analyzing the global properties of the solutions and for diagnosing potential spacetime singularities.

\subsection{Static solutions}

To obtain the static solution of the system, we need to solve the set of differential equations that characterize the static configuration. For this purpose, we impose boundary conditions at the horizon in the form of the following series expansion
\begin{align}
\zeta &= 1 + \frac{1}{2r_{h}}\left(\frac{Q^{2}(1+b\phi_{h}^2)}{ r_{h}^{2}} - 1\right)(r - r_{h}) + \cdots~,\label{eq:zeta1}\\
\phi &= \phi_{h} - \frac{2 b\phi_{h}^2Q^{2}}{2r_{h} \left[Q^{2}(1+b\phi_{h}^2) - r_{h}^{2} \right]}  (r - r_{h}) + \cdots~,\label{eq:phi1} 
\end{align}
where subscript $h$ represents the value at the horizon.

In asymptotically flat spacetime, the solution can be approximated as
\begin{align}
\zeta &= \sqrt{\frac{2M}{r}}\left(1 - \frac{Q^{2} + Q_{s}^{2}}{4Mr} + \cdots\right),\label{eq:zeta2}\\
\phi&=\frac{Q_{s}}{r}+\frac{MQ_{s}}{r^2}+\cdots~,\label{eq:phi2} 
\end{align}
where $Q_{s}$ represents the scalar charge. This means that for given $b$, $Q$, and $M$, the scalar charge $Q_{s}$ and $\phi_{h}$ can be determined. During the solving procedure, firstly, we fix  $b$, $Q$ and $r_{h}$, determine an initial trial value for $\phi_{h}$. The boundary conditions (\ref{eq:zeta1}, \ref{eq:phi1}) are enforced at $r_{b1}=r_{h}(1+\epsilon)$ with $\epsilon\approx 10^{-7}$. The coupled differential equations (\ref{eq:zeta}, \ref{eq:phi}) are then integrated numerically until $r_{b2}$ and $r_{b2}\approx10^6r_{h}$, if the field $\phi(r)$ asymptotically vanish at $r_{b2}$ according to (\ref{eq:phi2}), which means a static background solution is successfully identified. Conversely, if $\phi(r)$ remains finite at $r_{b2}$, the initial guess for $\phi_{h}$ must be refined, and the integration procedure iterated until convergence is achieved. Once the static background solutions $\zeta$ and $\phi$ are obtained, we can get the lapse function $\alpha$ from (\ref{eq:alpha}), and the total mass $M$ and the scalar charge $Q_{s}$ from (\ref{eq:M}) and (\ref{eq:phi2}), respectively.

\begin{figure}[htbp]
    \centering
    \begin{tikzpicture}
        \node[anchor=south west] at (0,0) {\includegraphics[height=6cm]{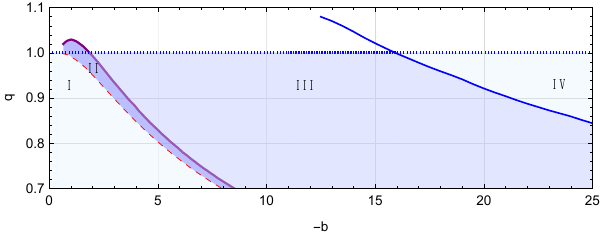}};
        \node[anchor=south west] at (3.0cm,4.7cm) {\includegraphics[width=4cm]{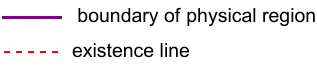}}; 
        \node[anchor=south west] at (10.0cm,5.1cm) {\includegraphics[width=4cm]{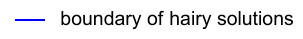}}; 
    \end{tikzpicture}
    \caption{Distribution of black hole solutions in the parameter space.
The red dashed curve denotes the boundary between hairy and hairless black hole solutions, also referred to as the existence line.
The purple solid curve marks the boundary of the physical region, corresponding to the divergence of the coupling function, which occurs when $1 + b \phi^2 = 0$, where $f(\phi)  $ becomes singular.
The blue solid curve indicates the boundary of scalar-hairy black hole solutions exhibiting negative energy density. For $q < 1$, the parameter space can be divided into four distinct regions according to the existence and physical properties of the black hole solutions:
Region~I lies below the existence line;
Region~II is located between the existence line and the physical boundary;
Region~III is the region above the physical boundary but below the scalar-hairy black hole boundary;
and Region~IV lies above the scalar-hairy black hole boundary.}
\label{fig:tikz-overlay}
\end{figure}

\begin{figure}[htbp]
    \centering
    \begin{tabular}{cc}
        \includegraphics[width=0.45\textwidth]{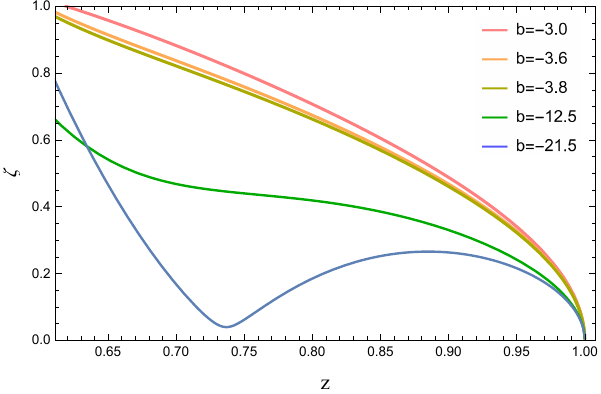} &
        \includegraphics[width=0.45\textwidth]{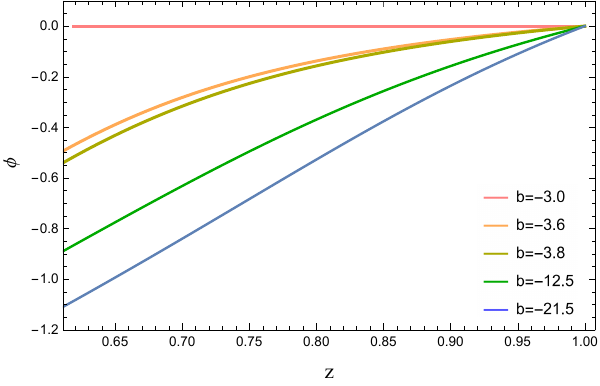} \\   
    \end{tabular}
    \caption{At a fixed charge parameter $q = 0.9$, The \textit{left panel} shows the metric function $\zeta(z)$ as a function of the radial coordinate $z$, while the \textit{right panel} displays the scalar field $\phi(z)$ as a function of $z$, where $z$ denotes the compactified radial coordinate. The five selected parameter sets are ordered by decreasing $b$ and are located, respectively, on the existence line separating bald and hairy solutions, in region~II, on the divergence line of the coupling function, in region~III, and on the existence boundary of hairy black hole solutions. As the parameters approach the existence boundary of the hairy solutions, the metric function $\zeta(z)$ exhibits a pronounced drop and tends toward zero at a finite radial position, while the radial profile of the scalar field $\phi(z)$ also undergoes significant changes, indicating a marked transition in the geometric and physical properties of the solutions in this region.} 
    \label{fig:combined7}
\end{figure}

In order to characterize the distribution of static black hole solutions, we introduce the dimensionless charge parameter $q \equiv Q/M $. The following analysis is carried out in the parameter space spanned by the coupling parameter $b$ and the charge parameter $q$.
After completing the above numerical computation\footnote{\href{https://github.com/valley-gorge/black-hole-dynamics-.git}{For further details, we refer the reader to the static analysis procedure}}, we obtained the distribution of static solutions in the parameter space, as shown in Fig.~\ref{fig:tikz-overlay}. The red dashed curve delineates the hairless black hole solutions from the hairy ones, while the solid purple curve marks the line of divergence for the coupling function\citep{Fernandes_2019}. Above this divergence line, the coupling function becomes negative, leading to hairy black hole solutions with negative energy.
However, previous analyses of the EMS model show that hairy black hole solutions above the divergence line do not extend indefinitely in parameter space\citep{Fernandes_2019}; instead, they terminate at a nontrivial boundary. In order to determine the location of this boundary, we employed a static shooting method to solve the field equations and gradually approached the limiting configuration by moving toward lager values of $-b$. Through this process, we identified the existence boundary of hairy black hole solutions above the divergence line, indicated by the blue solid curve in Fig.~\ref{fig:tikz-overlay}.
Near this boundary, static solutions can still be obtained, but the metric function exhibits a pronounced decrease at a finite radial position and approaches zero in that region. Once the parameters cross this boundary, the static shooting method no longer converges, indicating that no static hairy black hole solutions satisfying the boundary conditions exist beyond this line.
In order to further elucidate the physical significance of this boundary, we selected five representative parameter points at fixed $q = 0.9$, located on the existence line, in Region II, on the divergence line, in Region III, and on the boundary of hairy solutions. The corresponding metric functions are shown in  Fig.~\ref{fig:combined7}. A comparison of these representative cases reveals that the nature of the solutions changes significantly near the blue curve, and no static hairy black hole solutions exist beyond it. These analyses suggest that the blue curve not only marks the termination point of the solution family in the parameter space, but also likely corresponds to the critical boundary where the solutions transition from a regular structure to geometric degeneration.

Combining these results, we find that when $q >1$, only hairy black hole solutions exist. When $q<1$, the parameter space can be divided into four regions with distinct physical properties:
Region I: Only stable hairless black hole solutions exist.
Region II : Hairless and hairy solutions coexist. The hairy solutions are stable and possess positive energy outside the horizon, while the hairless RN black hole is linearly unstable.
Region III : The RN black hole solution persists but is unstable. Hairy black hole solutions can also be constructed, but they exhibit negative energy near the horizon\citep{Fernandes_2019}.
Region IV : Only unstable hairless RN black hole solutions exist.

\section{Analysis of the Dynamical Evolution Results}\label{S3}

Although the static analysis in Section~\ref{S2} revealed the  properties of static solutions in the parameter space, it cannot determine the dynamical fate of these solutions under perturbations. Two key questions remain to be addressed: (i) whether the hairless black holes in Regions II and III will, as predicted by linear stability analysis, ultimately evolve into the corresponding hairy solutions;and (ii) what the dynamical fate of the unstable RN solution in Region IV is once it is perturbed, how this evolution manifests itself within the fully nonlinear framework, and whether it may trigger pathological behavior in the spacetime structure.
In addition, we also examine the dynamical evolution of hairless solutions in Region I, which remain stable under perturbations, thereby providing further confirmation of the static stability picture. 

 In order to address these issues, in this section we perform fully nonlinear numerical simulations by perturbing the hairless solutions in all four regions\footnote{\href{https://github.com/valley-gorge/black-hole-dynamics-.git}{For further details, please refer to the dynamical evolution procedure}}. Our objectives are twofold: first, to verify the instability of hairless solutions in Regions II and III and to provide a comprehensive analysis and quantitative characterization of the stable hairy end states (such as the Kretschmann scalar and energy density distribution), thereby offering robust dynamical evidence for the stability diagram established in Section~\ref{S2}; Second, we explore whether the hairless black hole solutions in Region IV  may allow for violations of the WCCC during their nonlinear evolution.

We set an RN black hole  with mass $M_{0}=1$ and charge $Q=0.9$ as the initial seed black hole. Here $M_0$ sets the unit length scale of our simulation. 
Its metric function satisfies $\zeta_{0}(r)=\sqrt{\frac{2M_{0}}{r}-\frac{Q^{2}}{r^{2}}}$, and the background scalar field $\phi_{0}(r)=0$. The inner position of the computational domain is set to $r_{c}=1.3$ (in the vicinity of the initial horizon). We introduce $z = r/(1 + r/M_0)$ to compactify the coordinates. The system is evolved within the domain $z \in [z_{c}, 1)$, where $z=1$ corresponds to spatial infinity and $z_{c} = r_{c}/(1 + r_{c})=0.5652$. We uniformly discretize the $z$-coordinate into $2^{12}$ grid points, employing the finite difference method in the radial direction and the fourth-order Runge-Kutta method for temporal evolution.

At the initial time, we introduce an ingoing scalar perturbation wave as follows:
\begin{align}
\delta\phi(r) &= p\left\{
\begin{array}{ll}
e^{-\frac{1}{r-r_{1}}-\frac{1}{r_{2}-r}}(r_{2}-r)^{2}(r-r_{1})^{2},~~ & r_{1} < r < r_{2}~, \\
0,~~ & \text{otherwise}~,
\end{array}
\right. \label{eq:delta_phi} \\
\Pi(r) &= \partial_{r}\delta\phi(r) \label{eq:Pi_definition}~.
\end{align}
Here, $p$ represents the amplitude and is set to $10^{-3}$, with $r_1=4$ and $r_2=9$.
Given the initial $\phi$, $\Pi$ and $\Phi=\partial_r \phi$, the shift function $\zeta$ can be obtained from  (\ref{eq132}) by applying Newton-Raphson method with boundary conditions as $\left.\zeta\right|_{t = 0, r = r_{c}} = \zeta_{0}(r_{c})$. Next, using (\ref{eq131}), we obtain the initial spatial distribution of $\alpha$. After determining the initial values of $\Phi$, $\Pi$, $\zeta$, and $\alpha$, we evolve $\zeta$, $\phi$, and $\Pi$ to the next time step by solving the evolution equations \eqref{eq133}-\eqref{eq142}. The values of $\Phi$ and $\alpha$ can be determined from the auxiliary relations (\ref{eq122}) and (\ref{eq131}), respectively. Repeating this iterative process allows us to calculate the metric and scalar field functions at each time step.  Since matter cannot reach spatial infinity within finite time,  we impose the boundary conditions at spatial infinity:  $\left.\alpha\right|_{r \to \infty} = 1$ and $\left.\phi,\Pi,\Phi\right|_{r \to \infty} = 0$, which means $\partial_{t}\phi|_{r \to \infty}=\partial_{t}\Pi|_{r \to \infty}=0$. Note that the constraint  (\ref{eq132}) is only required once during the initial setup.

\begin{figure}[htbp]
    \centering
    \begin{tabular}{cc}
        \includegraphics[width=0.45\textwidth]{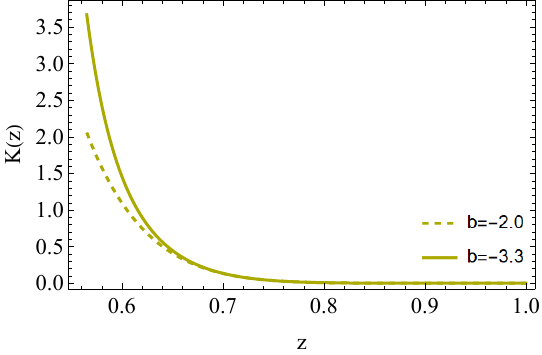} &
        \includegraphics[width=0.46\textwidth]{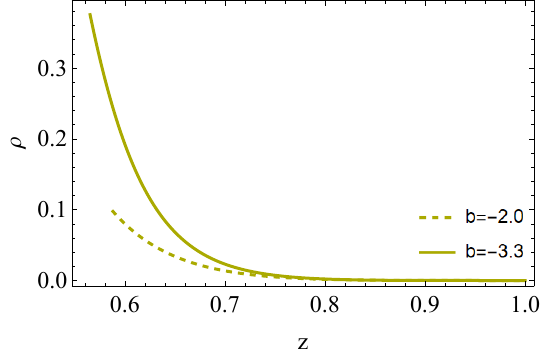} \\   
        \includegraphics[width=0.46\textwidth]{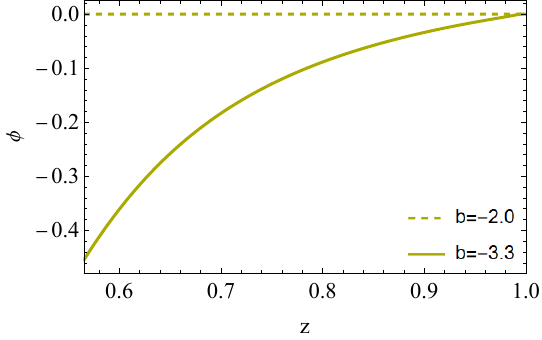} &
        \includegraphics[width=0.45\textwidth]{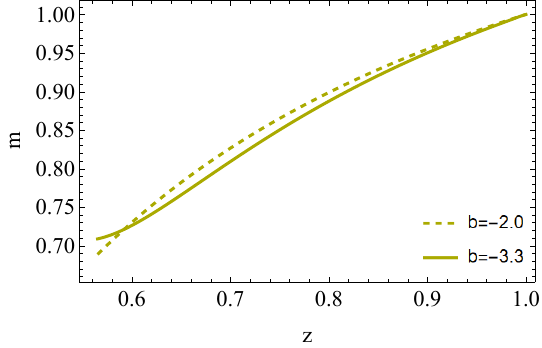} \\  
    \end{tabular}
    \caption{ The dashed curve represents the RN solution in Region I, while the solid curve corresponds to the hairy solution in Region II.  Both classes of solutions are physically regular. The\textit{ top-left panel}  shows the Kretschmann scalar, which remains finite from the immediate vicinity of the horizon to spatial infinity, indicating a well-behaved spacetime geometry. The\textit{ top-right panel} shows the spatial distribution of the energy density, exhibiting a monotonically decreasing trend. The\textit{ bottom-left panel} illustrates the spatial profile of the scalar field, which increases monotonically. The\textit{ bottom-right panel} presents the Misner-Sharp mass, growing monotonically from near the horizon to infinity, suggesting an energy distribution extending to infinity. }
    \label{fig:combined1}
\end{figure}
\begin{figure}[htbp]
    \centering
    \begin{tabular}{cc}
        \includegraphics[width=0.45\textwidth]{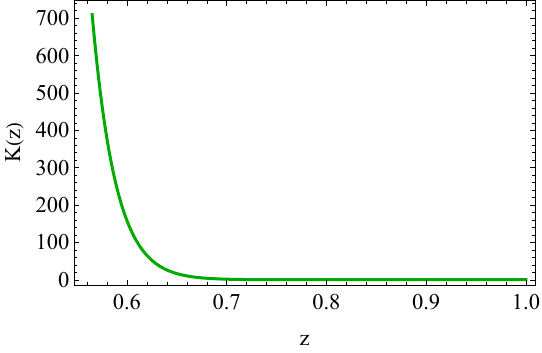}&
        \includegraphics[width=0.46\textwidth]{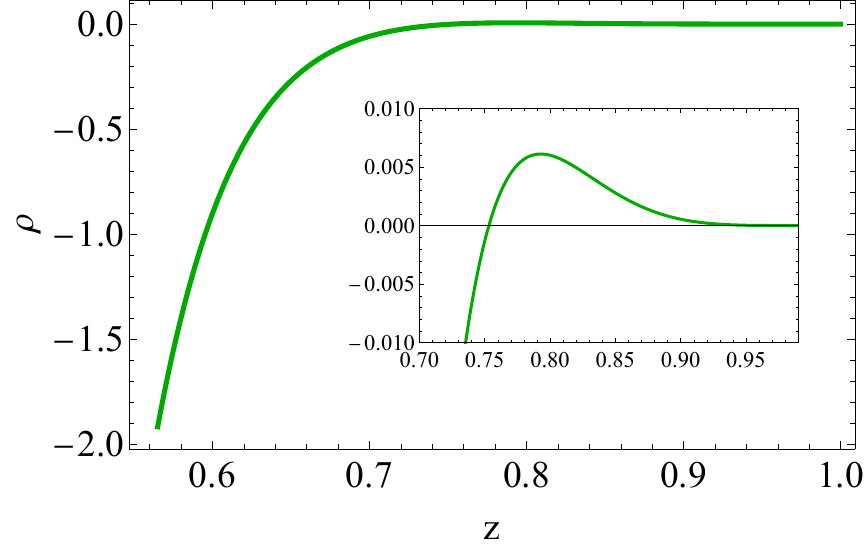}\\
        \includegraphics[width=0.47\textwidth]{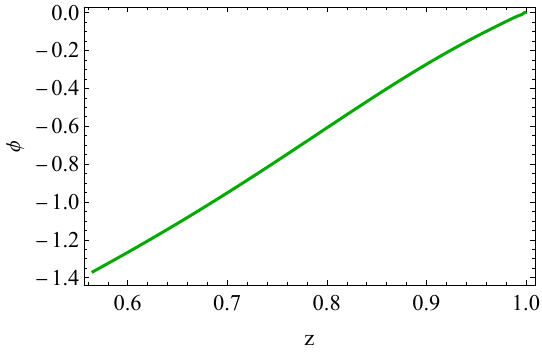} &
        \includegraphics[width=0.45\textwidth]{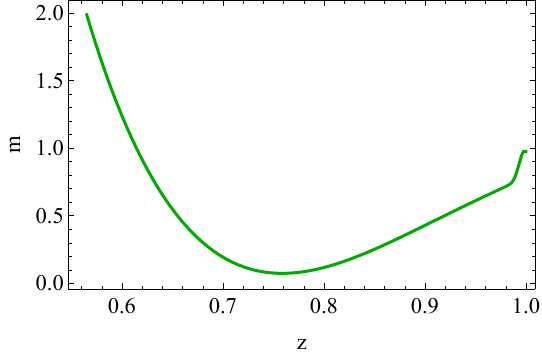} \\   
    \end{tabular}
  
    \caption{ The figure corresponds to the parameter choice $b=-10$. The \textit{top-left panel} shows the Kretschmann scalar. The \textit{top-right panel} shows the energy density. The inset shows a magnified view of the zero-crossing region. The \textit{bottom-left panel} and \textit{bottom-right panel} respectively represents the spatial distribution of the scalar field and Misner-Sharp mass. It is interesting that in the region with negative energy density, the Misner-Sharpe mass shows a decreasing trend. On the contrary, once the energy density becomes positive, it begins to increase.}
    \label{fig:combined2}
\end{figure}

\subsection{Final States of Perturbed Black Holes}

By employing the numerical methods presented above, we performed nonlinear evolutions of the hairless black hole seeds in all three parameter regions of Fig.~\ref{fig:tikz-overlay} and obtained their characteristic final-state structures. Fig. \ref{fig:combined1} illustrates the typical final-state structures in Region I  and Region II. In these two regions, the Kretschmann scalar, energy density, MS mass, and scalar field distribution outside the black hole horizon all remain regular, and the energy density is consistently positive, thereby satisfying the weak energy condition. In contrast, Fig. \ref{fig:combined2} presents representative results for Region III, where the energy density exhibits a pronounced negative value near the horizon and gradually becomes positive as the radial coordinate increases. Although the Kretschmann scalar, scalar field, and MS mass remain regular, the occurrence of negative energy density is an anomalous phenomenon, indicating a violation of the weak energy condition in the vicinity of the horizon and suggesting that the WCCC may be disrupted in this region. The distribution of the MS mass further supports this possibility.

We conjecture that at parameter points farther away from the divergence line in the parameter space of Fig.~\ref{fig:tikz-overlay}, the negative-energy effect will become more significant. In order to test this conjecture and further study the resulting spacetime structure, we extend our simulations to larger parameter values, setting $b=-100$.

\begin{figure}[htbp]
    \centering
    \begin{tabular}{cc}
        \includegraphics[width=0.46\textwidth]{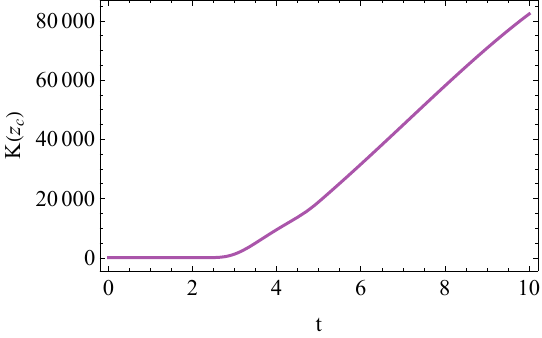} &
        \includegraphics[width=0.47\textwidth]{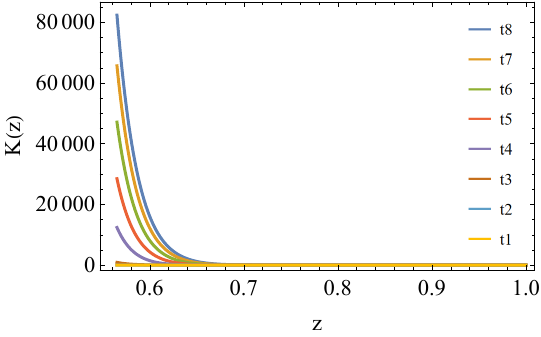} \\  
        \includegraphics[width=0.45\textwidth]{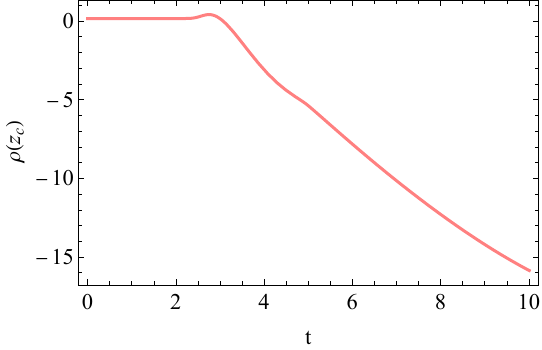} &
        \includegraphics[width=0.46\textwidth]{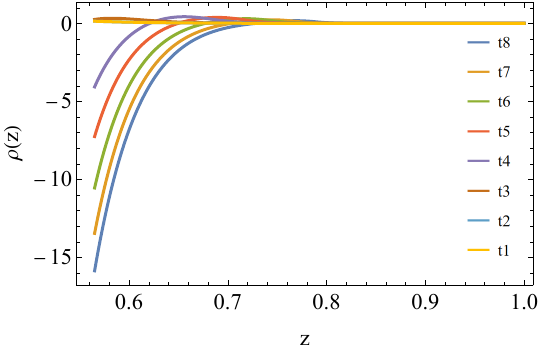} \\ 
    \end{tabular}
    \caption{Spacetime evolution of the Kretschmann invariant and the energy density for $b=-100$.  \textit{top-left panel}: the temporal evolution of $K(z_{c})$ over $t\in[0,10]$. Its steady growth reflects the continuous enhancement of spacetime curvature. \textit{top-right panel}:  the spatial profiles of $K(z)$ in the radial interval $z\in[z_c,1)$ at evolution times $t_1$ to $t_8$ (here, $t_1 \approx 0.104$, and $\Delta t\approx 1.415$). The curvature peak continuously increases, indicating rapid amplification of curvature and the potential violation of WCCC.  \textit{bottom-left panel}: the temporal evolution of $\rho(z_{c})$ over $t\in[0,10]$. In the later stages, the curve shows a persistent downward trend and reaches significantly negative values, further confirming the breakdown of energy conditions in this region. \textit{bottom-right panel}: the spatial profiles of the energy density $\rho(z)$ in the radial interval  $z\in[z_c,1)$ at evolution times $t_1$ to $t_8$. The energy density becomes increasingly negative, revealing a deepening violation of the energy conditions.}  
    \label{fig:combined4}
\end{figure}

\subsection{Potential violation of WCCC}

In this subsection, we present numerical simulation results showing that when the coupling parameter $b$ takes a specific value within Region IV (namely, $b = -100$), the dynamical evolution of the system may lead to the formation of a naked singularity. We begin by examining the evolution of the Kretschmann scalar $K$. As shown in the upper-left panel of Fig.~\ref{fig:combined4}, at a fixed spatial point inside the event horizon, $K$ increases monotonically and rapidly with time. This behavior indicates a sharp growth of the local spacetime curvature and suggests an approach toward curvature divergence. To assess whether this singular behavior is confined to the black hole interior or extends beyond the horizon, we further analyze the spatial distribution of $K$ at different evolution times, as shown in the upper-right panel of Fig.~\ref{fig:combined4}. The results demonstrate that the values of $K$ in the vicinity of the event horizon also grow significantly as the evolution proceeds. This observation suggests that the curvature divergence is not restricted to the black hole interior but may propagate toward, or even outside, the horizon. We note that the numerical evolution inevitably breaks down once $K$ becomes sufficiently large, preventing us from directly resolving the final formation of a naked singularity. Nevertheless, the observed growth trend of $K$ near and outside the horizon strongly indicates the development of singular features visible to external observers, thereby signaling a potential violation of the weak cosmic censorship conjecture.

To clarify the physical origin of the curvature growth discussed above, we examine the time evolution of the energy density in the vicinity of the black hole. As shown in the lower panels of Fig.~\ref{fig:combined4}, at a representative point inside the event horizon, the energy density becomes negative at an early stage of the evolution and subsequently decreases in magnitude.
The spatial profiles further indicate that a similar behavior develops in the neighborhood of the outer horizon at later times, suggesting that the region characterized by negative energy density is not restricted to the deep interior but gradually extends toward the horizon. The condition $\rho<0$ corresponds to a violation of the weak energy condition, which is known to invalidate several key assumptions underlying classical horizon stability arguments.
In this context, the appearance of negative energy density provides a consistent physical interpretation of the observed growth of curvature invariants. While the numerical evolution cannot resolve the ultimate end state once the curvature becomes sufficiently large, the correlated emergence of energy condition violations and rapidly increasing curvature supports the interpretation that the system is dynamically driven toward horizon instability, with possible implications for the validity of the weak cosmic censorship conjecture.

\begin{figure}[htbp]
    \centering
    \begin{tabular}{cc}
        \includegraphics[width=0.48\textwidth]{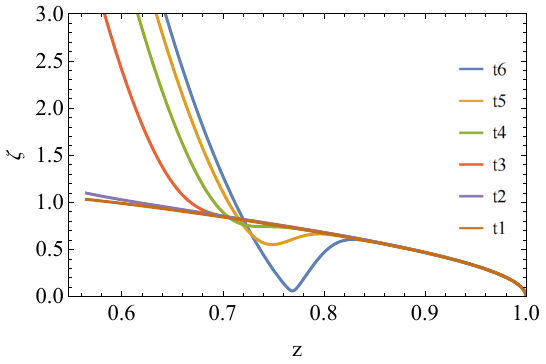} &
        \includegraphics[width=0.46\textwidth]{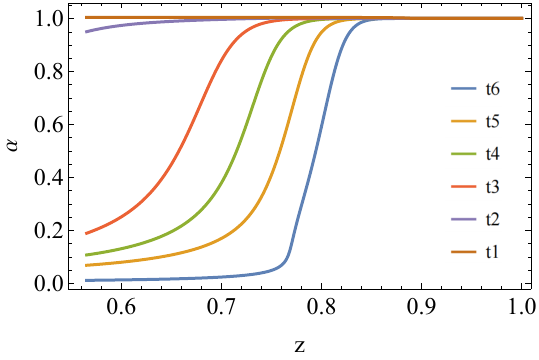} \\  
    \end{tabular}
    \caption{The \textit{left panel} shows the time evolution of the shift function $\zeta$, with different colors indicating different evolution times ($t_1 \approx 7\times10^{-4}$, and $\Delta t\approx 2.3817$). From the early to the late stages of the evolution, the curve gradually changes from a gentle decline to a V-shaped structure characterized by a decrease followed by a rise. In the final stage, the curve reaches a minimum near $z\approx 0.78$, and then exhibits a slight rebound around $z\approx 0.82$. As $z\to1$, all curves approach zero, indicating that the spacetime becomes asymptotically flat and the shift function tends to vanish.
    The \textit{right panel} shows the radial evolution of the lapse function $\alpha(z)$. In the PG coordinates, $\alpha$ characterizes the ratio between the coordinate time and the proper time of a timelike observer, with $\alpha<1$ in strong-gravity regions. At the early stage of the evolution, the curves exhibit smooth S-shaped profiles and gradually shift toward larger $z$. At the final time slice $(t_{6})$, a pronounced dip emerges around $z\approx0.8$, marking the formation of a strong-gravity region.}\label{fig:combined5}
\end{figure}

In addition to the pronounced growth of the Kretschmann invariant $K$  and the emergence of negative energy density discussed above, we further examine several geometric quantities that are directly tied to the spacetime structure in order to better characterize the degenerative behavior observed during the evolution. In particular, we focus on the shift function 
$\zeta$ and the lapse function 
$\alpha$, whose time evolution exhibits clear qualitative changes. Based on their behavior, the entire dynamical process can be naturally divided into three stages: early, intermediate, and late, as illustrated in Fig.~\ref{fig:combined5}.

During the early stage, the evolution of the geometric variables remains consistent with that of a perturbed RN exterior solution. Both 
$\alpha$ and $\zeta$ vary smoothly in time and do not display any anomalous features, indicating that the spacetime geometry remains close to its initial configuration. As the system enters the intermediate stage, the lapse function $\alpha$ begins to decrease rapidly in the vicinity of the horizon, while the shift function 
$\zeta$ grows significantly in the strong-field region. These changes signal the onset of nonlinear amplification of the geometric variables near the horizon and the gradual development of a strongly curved regime.

The most pronounced nonlinear behavior occurs in the late stage of the evolution. Shortly before the numerical integration terminates, the lapse function $\alpha$ undergoes a rapid collapse, while the shift function $\zeta$ exhibits explosive growth near the central region. We emphasize that this correlated divergence of geometric quantities has been consistently observed across multiple independent numerical implementations, including simulations based on the BSSN formulation. This robustness strongly suggests that the observed behavior reflects an intrinsic instability of the spacetime geometry, rather than numerical artifacts associated with specific gauge choices.

Overall, although the present simulations do not directly resolve the formation of a naked singularity, the divergence of $\zeta$ and the collapse of $\alpha$, together with the appearance of negative energy density in the vicinity of the outer horizon, points toward a dynamical weakening of the trapping structures under strong nonlinear evolution. If this trend persists over longer evolution times, the strongly curved region may, in principle, approach a location that is no longer fully shielded by a stable event horizon. In this sense, while our results do not constitute a definitive counterexample to the weak cosmic censorship conjecture \citep{Penrose:1969pc,wald1997gravitationalcollapsecosmiccensorship}, they reveal a class of dynamical evolutions within the present framework for which a clear verification of the conjecture becomes challenging.

\section{Discussion}\label{S4}

In this work, we investigated both the static properties and the nonlinear dynamical evolution of black holes in the Einstein–Maxwell–Scalar theory with fractional coupling. Our static analysis demonstrates that, within a finite region of parameter space, scalarized black hole solutions develop negative energy density in the vicinity of horizon. This feature signals a violation of the classical energy conditions and suggests that the horizon-supporting structure may already be intrinsically weakened at equilibrium.

The dynamical simulations presented here provide further support for this interpretation. For moderate values of the coupling parameter, perturbed RN black holes relax toward regular scalarized configurations, consistent with previous studies. In contrast, when the fractional coupling becomes sufficiently strong (e.g., $b=-100$), the spacetime evolution exhibits qualitatively different behavior. In this regime, curvature invariants such as the Kretschmann scalar grow rapidly and monotonically, while the lapse function 
$\alpha$ and shift function $\zeta$ undergo synchronized nonlinear deformation. At the same time, negative energy density near the outer horizon persists and deepens throughout the evolution. The coherent development of these features in the strong-field region indicates a progressive weakening of the trapping structure during the nonlinear evolution.

Although the present simulations do not directly resolve the complete destruction of the event horizon or the explicit formation of a naked singularity, the combined evidence from curvature growth, geometric degeneration, and sustained violations of the weak energy condition points toward a dynamical trend consistent with horizon destabilization in the strong fractional-coupling regime. Importantly, these behaviors have been consistently observed across multiple independent numerical implementations, including simulations based on the BSSN formulation, indicating that the effect is robust and not attributable to a specific gauge choice or numerical scheme.

From a broader perspective, our results differ in an essential way from previous studies of potential cosmic censorship violation in higher-curvature gravity theories \citep{Corelli:2022phw,Corelli:2022pio}, where the breakdown of hyperbolicity or formulation-dependent pathologies complicate the physical interpretation. In the present case, the evolution is governed by a well-posed hyperbolic system in an asymptotically flat spacetime, and the observed instabilities are directly correlated with local violations of classical energy conditions induced by the fractional coupling. This suggests that the mechanism driving the weakening of the horizon is physical in origin, rather than an artifact of the underlying formulation.

Nevertheless, fully determining the ultimate fate of the spacetime requires further investigation. The obstruction encountered here is not numerical or formulation-dependent, but reflects the rapid development of strong curvature that prevents a direct assessment of the global horizon structure. While local diagnostics robustly indicate horizon failure, confirming the global structure of the resulting singularity remains an open challenge. 
Extending the analysis to rotating configurations may further illuminate the generality of the mechanism identified here and its implications for cosmic censorship in Einstein–Maxwell–Scalar theories.

\section{Acknowledgments}

This work is supported by the National Natural Science Foundation of China (Grants No. 12575055, 12247101 and 12375048) and "Talent Scientific Fund of Lanzhou University".

\bibliographystyle{modified-apsrev4-2}  
\bibliography{refs}

\end{document}